\documentclass{jfm}
\usepackage{color}
\usepackage{amstext}
\usepackage{amssymb}
\usepackage{graphicx}
\usepackage{amsmath}
\usepackage{bm}
\usepackage[mathscr]{euscript}
\usepackage{amsfonts}
\usepackage{caption}

\usepackage{tcolorbox}
\newtcolorbox{mybox}[1]{colback=red!5!white,colframe=red!75!black,fonttitle=\bfseries,title=#1}

\newcommand\be{\begin{equation}\,}
\newcommand\ee{\end{equation}\,}

  \newcommand{\thalf}{\textstyle{\frac{1}{2}} \displaystyle }

\makeatletter
\def\@makefnmark{\hbox{\@textsuperscript{\normalfont\@thefnmark}}}
\makeatother

\begin{document}

{\title[Chiral transfer of angular momentum]
{Chiral transfer of angular momentum}}

\author[H. K. Moffatt and V. A. Vladimirov ]
{H. K. Moffatt$^1$ and V. A. Vladimirov$^{2,3,4}$}

\affiliation{$^1$Department of Applied Mathematics and Theoretical Physics, \\
Wilberforce Road, Cambridge CB3 0WA, UK\\
[\affilskip]$^2$Sultan Qaboos University, Oman,\ $^3$University of York, UK,\ $^4$University of Leeds, UK}

%\affiliation{$^1$University of Cambridge, UK, \
%       $^2$Sultan Qaboos University, Oman and University of York, UK}

%\author[V. A. Vladimirov]
%{V.\ns A.\ns V\ls l\ls a\ls d\ls i\ls m\ls i\ls r\ls o\ls v$^{1,2,3}$}

\setcounter{page}{1}\maketitle \thispagestyle{empty}

\begin{abstract}
\noindent Suppose that viscous fluid is contained in the space between a fixed sphere $S_2$  and an interior sphere $S_1$ which moves with time-periodic velocity ${\bf U}(t)$ and angular velocity
${\bf \Omega}(t)$, with $ \left<{\bf U}(t)\right> = \left<{\bf \Omega}(t)\right> = 0$.   It is shown that, provided this motion is chiral in character, it can drive a flow that exerts a non-zero torque on $S_2$.  Thus angular momentum can be transferred through this mechanism.
\end{abstract}

\section{Introduction}
Consider the following simple problem:  suppose that a sphere $S_1$ of radius $r_1$  is contained inside a fixed sphere $S_2$ of radius $r_2 >r_1$, the space between being filled with viscous fluid.
Suppose that $S_1$ is moved with a time-periodic velocity ${\bf U}(t)$ and angular velocity ${\bf\Omega} (t)$ with zero time average:  $\left<{\bf U}(t)\right>=\left<{\bf\Omega} (t)\right>=0$, as in the sketch of Figure \ref{Fig_config}(a).  Is it possible that such a motion can generate a mean torque on the fixed sphere $S_2$?  We shall show by explicit example that the answer is positive, the effect arising only if the flow between the spheres has the property of chirality (lack of reflection symmetry); in this respect it is analogous to the helicity effect that is responsible for the self-excitation of magnetic field in a conducting fluid (Moffatt 1978).

\section{A simple planar model}\label{Sec_cartesian}
\begin{figure}
        \centering
        \begin{minipage}{.5\textwidth}
            \centering
            \includegraphics[width=.7\linewidth]{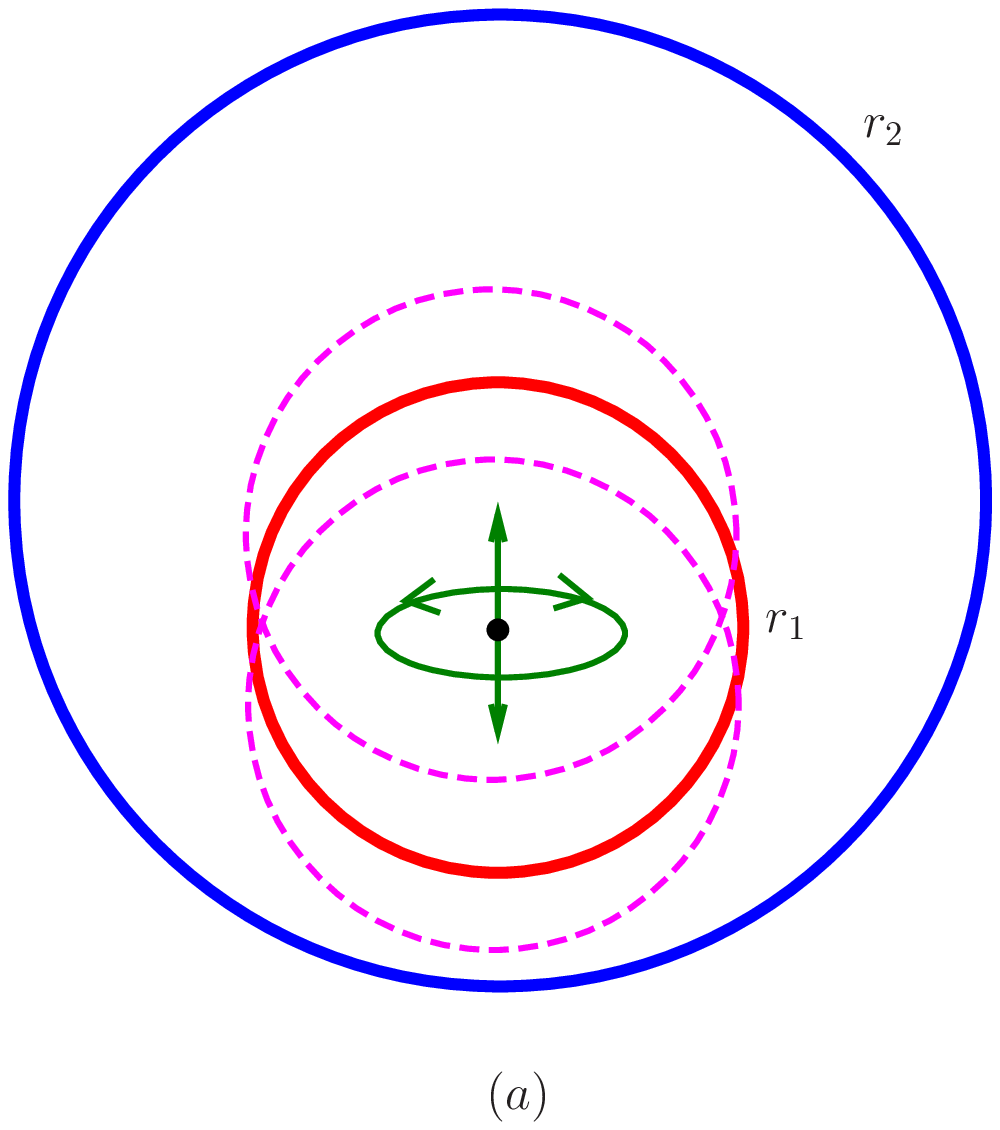}
        \end{minipage}%
        \begin{minipage}{.5\textwidth}
            \centering
            \includegraphics[width=.9\linewidth]{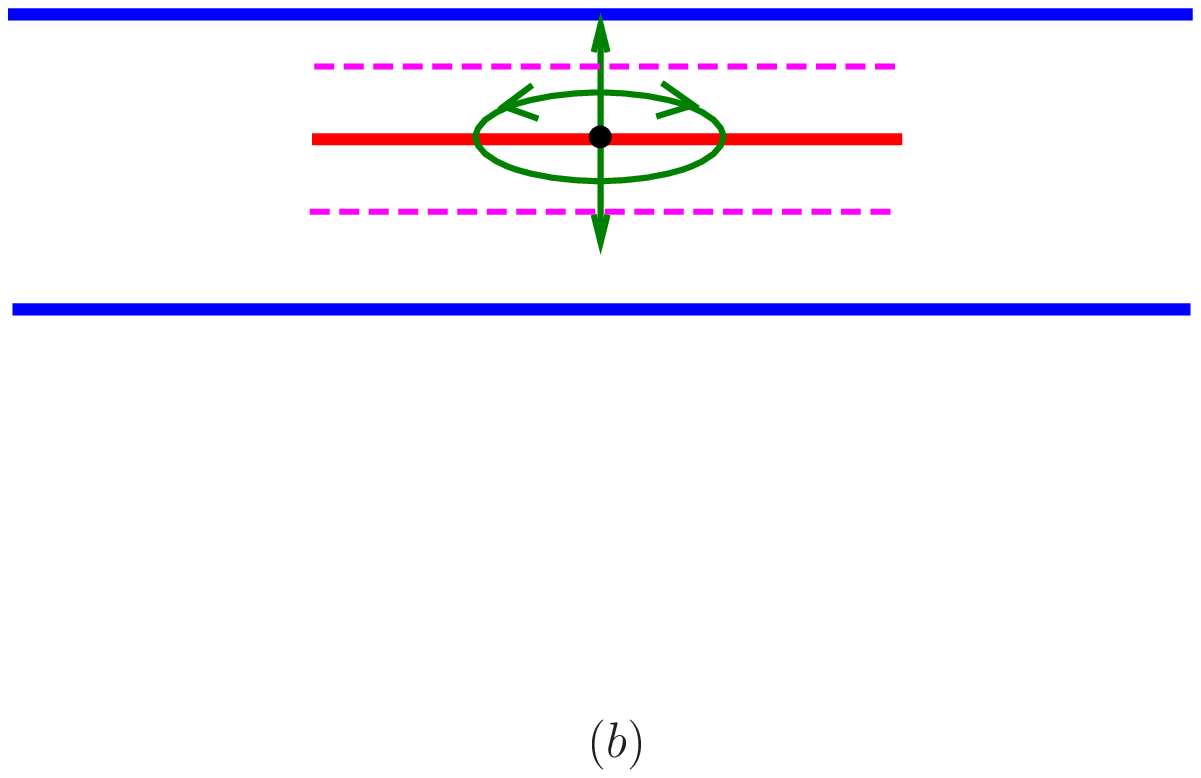}
        \end{minipage}
        \caption{ (a) Two-sphere configuration; the outer sphere is fixed and the inner sphere oscillates between the (dashed) limits indicated, and rotates about the axis of symmetry with time-periodic angular velocity $\omega(t)$ with $\left<\omega(t)\right>=0$. (b) Corresponding oscillating-disc model.}
        \label{Fig_config}
    \end{figure}

Suppose first that  fluid of viscosity $\mu$ and kinematic viscosity $\nu = \mu/\rho$ fills the space $|z|<1$ between two rigid plane boundaries $z=\pm 1$, and  that a circular disc of radius $a\gg 1$ and negligible thickness is immersed in the fluid parallel to the boundaries at position  $z=d_{0}$ with 
$|d_{0}|<1$.  We adopt cylindrical polar coordinates $(r,\varphi, z)$, where $r$ is the radial distance from the axis of the disc. The situation is sketched in Figure \ref{Fig_config}(b).
%\vskip 2mm
%
%\textcolor[rgb]{1.00,0.00,0.00}{Fig. 1a,b from the Draft 8:  the general view of spheres and disc}
%
%\vskip 2mm
Suppose now that the disc is caused to oscillate vertically so that its position  at time $t$ is $d(\tau)=d_0+\zeta(\tau)$, where $\tau\equiv \sigma t$ and $\sigma$ is a frequency. The disc is simultaneously caused to rotate about its axis with angular velocity $\omega(\tau)$. Both functions $\zeta(\tau)$ and $\omega(\tau)$ are assumed to be $2\pi$-periodic with zero average, $\langle \zeta(\tau)\rangle=\langle \omega(\tau)\rangle=0$.

Let $U_0$ be the maximum speed of the disc during its periodic motion. We assume that the Reynolds number Re $=U_0 a/\nu$ and the Stokes number St $ = \sigma a^2/\nu $ are small, so that inertia effects in the fluid may be neglected;  the methods of `thin-film' lubrication theory (e.g. Batchelor 1967) are then applicable.
In these circumstances, the poloidal $(r,z)$ and toroidal ($\varphi$) components of the equations of motion are decoupled. We need here consider only the toroidal component of velocity $v (r,z,\tau)$ within the gaps $-1<z<d$ and $d<z<1$.  This is essentially a `Couette-flow' situation in both gaps,  with boundary conditions $v=0$ at $z=\pm 1$ and $v=\omega \,r$ at $z=d \pm 0$. A simple calculation of the total instantaneous torque  $G(\tau)$ exerted on the plates yields
\begin{eqnarray}\label{G}
G(\tau)= \kappa\,\omega\, D\quad \textnormal{where} \quad D\equiv \frac{1}{2}\left(\frac{1}{1-d}+\frac{1}{1+d}\right)= \frac{1}{1-d^2}\quad\textnormal{and}\quad\kappa\equiv \pi\mu a^4.
\end{eqnarray}
Hence the average torque on the plates over a period of the disc motion is
\begin{eqnarray}\label{G1}
\langle G(\tau)\rangle=\kappa \,\chi\quad\textnormal{where}\quad \chi\equiv\langle\omega(\tau) D(\tau)\rangle.
\end{eqnarray}
\!(There is an equal and opposite mean torque exerted by the fluid on the moving disc.)
\begin{figure}
        \centering
         (a)\begin{minipage}{.30\textwidth}
            \centering
           \includegraphics[width=.9\linewidth]{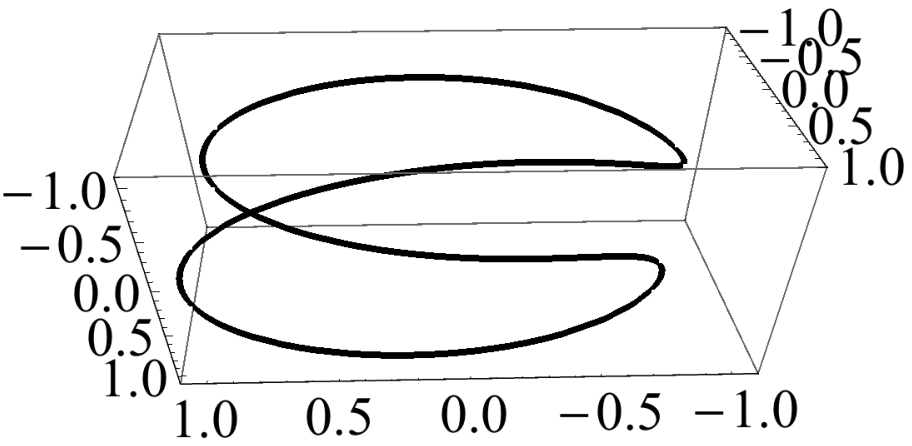}
        \end{minipage}%
        (b) \begin{minipage}{.30\textwidth}
            \centering
             \includegraphics[width=.9\linewidth]{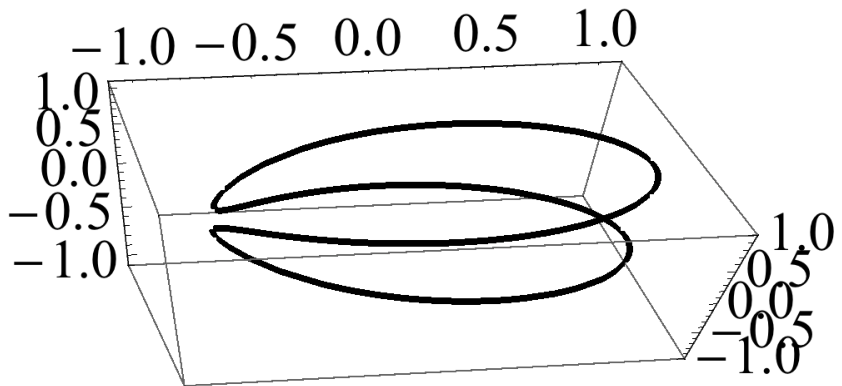}
        \end{minipage}
          (c) \begin{minipage}{.27\textwidth}
            \centering
           \includegraphics[width=.9\linewidth]{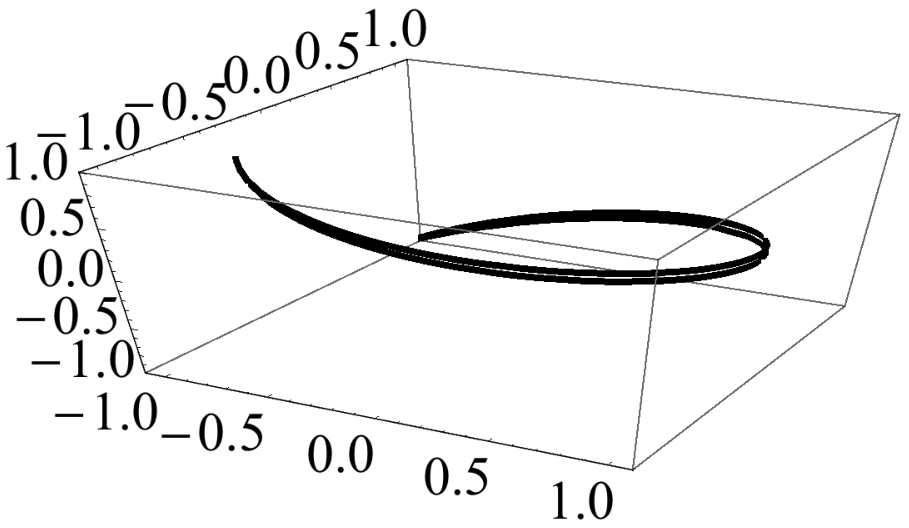}
        \end{minipage}
     \caption{Trajectories of a point on the disc with velocity (\ref{reg1}); $r_0=1, d_0 = 0.2, 
\lambda=0.76, \omega_0 =2.5$; (a) $\phi_0=0$, (b) $\phi_0 =\pi/4$, (c) $\phi_0=\pi/2 - 0.05$. }
\label{Fig_01_traj}
    \end{figure} 
\begin{figure}
        \centering
    (a)\begin{minipage}{.29\textwidth}
            \centering
           \includegraphics[width=.9\linewidth]{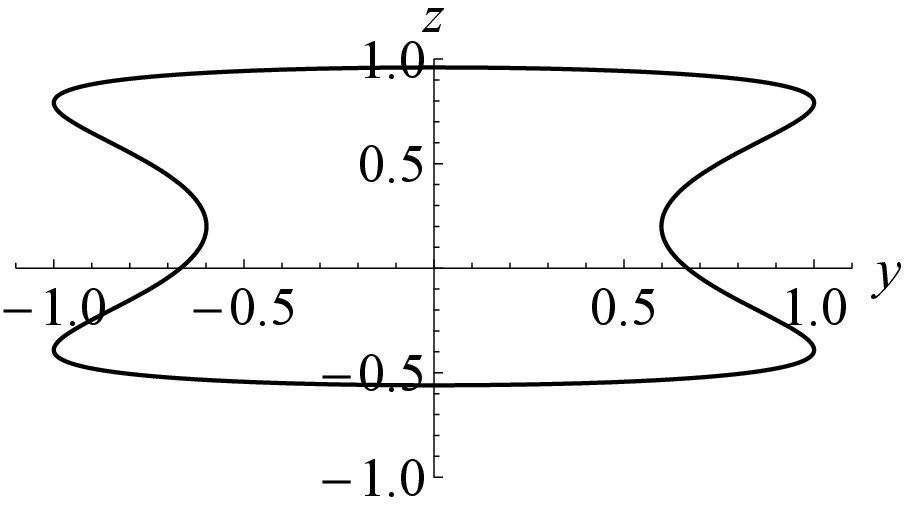}
        \end{minipage}%
        (b)\begin{minipage}{.29\textwidth}
            \centering
              \includegraphics[width=.9\linewidth]{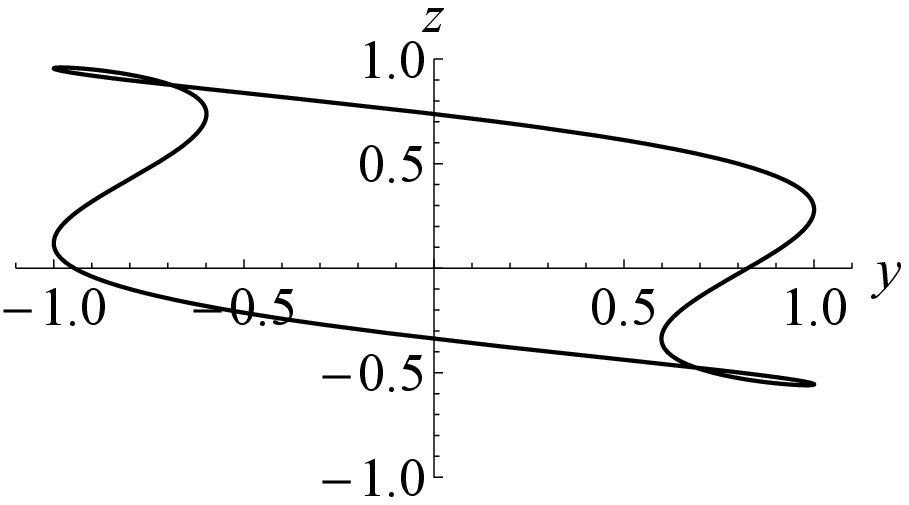}
       \end{minipage}
         (c)\begin{minipage}{.29\textwidth}
            \centering
              \includegraphics[width=.9\linewidth]{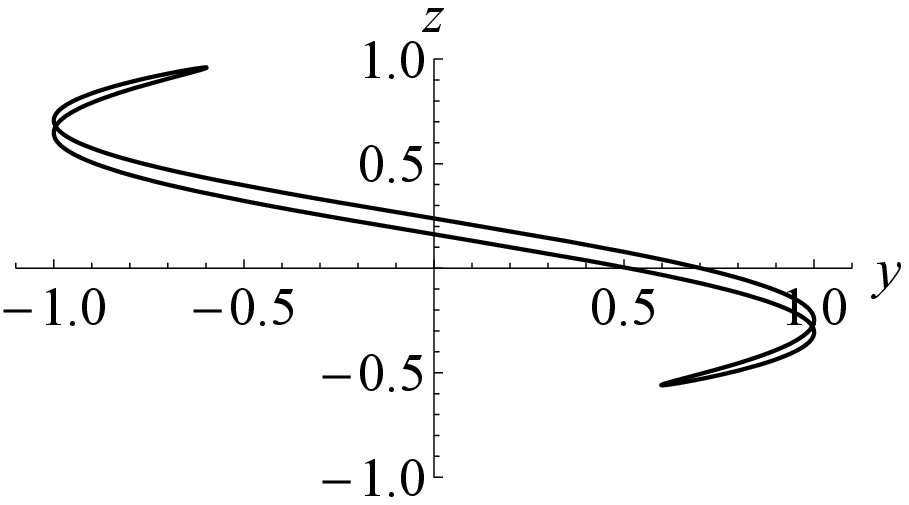}
        \end{minipage}
\caption{Projections on the plane $x=0$ of the trajectories of Figure \ref{Fig_01_traj}. }
\label{Fig_01_proj}
    \end{figure}
   
By way of example, let
\begin{eqnarray}\label{reg1}
 d(\tau)=d_0+\zeta(\tau),\quad \zeta(\tau)=\zeta_0 \cos \tau,\quad \omega(\tau)=\omega_{0}\cos(\tau+\phi_0),
\end{eqnarray}
where $|d_0 \pm \zeta_0|<1$ and where $\omega_{0}, \phi_0, d_0,  \zeta_{0}$ are non-negative constants. 
\!The trajectory of any point of the disc then has parametric equation $r\!=\!r_{0},\,\varphi\!=\varphi_0+\!\int_{0}^{\tau} \omega (\tau')\,\textnormal{d}\tau', \,z\!=\!d(\tau)$, and, being periodic in $\tau$ with period $2\pi$, is a closed curve on the cylinder $r=r_{0}$. Three examples of such trajectories, with $r_0=1, \,d_0=0.2, \,\zeta_0=0.76,\, \omega_0 =2.5$, and phase differences  $\phi_0=0,\pi/4$ and $\pi/2-0.05$, are shown in Figure \ref{Fig_01_traj}. These trajectories wind almost once  round the cylinder $r\!=\!r_{0}$, then reverse and wind back so that the closed curve makes zero net turns round the cylinder, as most clearly seen in Figure \ref{Fig_01_traj}(c). For the limiting case  $\phi_0=\pi/2$, the trajectory is a portion of a helix traversed up and down periodically.  Figure \ref{Fig_01_proj} shows projections of the same three figures on the plane $x=0$; note that this projection encloses a finite area (with positive or negative contributions from regions whose boundaries are traversed in an anticlockwise or clockwise sense) which decreases (actually proportional to $\cos \phi_0$) as $\phi_0$ increases from $0$ to $\pi/2$. 

Evaluation of the mean $\langle G\rangle$ from \eqref{reg1} now yields:
\begin{eqnarray}\label{mean_torque_1}
 \langle G\rangle=\kappa\,\omega_{0}\, g(d_0,\zeta_0)\cos \phi_0,
 \end{eqnarray}
 \noindent where
\begin{eqnarray}\label{mean_torque_11}
 g(d_0,\zeta_0)\equiv \frac{1}{2\zeta_0} \left[\frac{1-d_0}{\sqrt{(1-d_0)^2-\zeta_0^2}}-\frac{1+d_0}{\sqrt{(1+d_0)^2-\zeta_0^2}}\right]\,.
\end{eqnarray}
\begin{figure}
\vskip 5mm
\begin{center}
\includegraphics[width=0.50\textwidth]{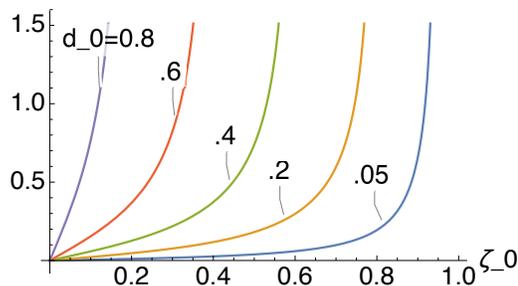}
\hspace*{3pt}
\end{center}
\caption{ The functions $g(d_0,\zeta_0)$ defined by (\ref{mean_torque_11}) for five values of $d_{0}$ and for $0<\zeta_0<1-d_0$ in each case.}
\label{Fig_G}
\end{figure}
\!\!\!\!\!\! The functions $g(d_0,\zeta_0)$  are shown in Figure \ref{Fig_G} for $d_{0}=0.2,0.4,0.6,0.8$ and the relevant range in each case, $0\!<\!\zeta_0\!<\!1-d_0$. Note the singular behaviour as $\zeta_0\rightarrow 1-d_0$; this simply reflects the torque singularity that is to be expected when the gap between the disc and the boundary $z=1$ tends to zero.

It is easy to understand the physical origin of this mean torque. Consider just the time interval $-\pi<\tau\le \pi$. If $d_0>0$ and $\zeta_0>0$, then the disc is at its shortest distance from the upper fixed boundary when $\tau = 0$; if $\phi_0 = 0$, the instantaneous torque on this boundary is then maximal and in the positive  direction (i.e.~the direction of increasing $\varphi$); when $\tau=\pi$, the disc is at its furthest distance from the upper boundary (as most evident in Figure \ref{Fig_01_proj}(a)) and the instantaneous torque is then minimal and in the negative direction.  Hence, the averaged torque on the upper boundary is positive. At the same time, a similar consideration for the more remote lower boundary results in a smaller average torque in the negative direction. The time-averaged total torque is therefore positive when $\phi_0=0$, as confirmed by (\ref{mean_torque_1}) and (\ref{mean_torque_11}) and Figure \ref{Fig_G}. Note the $\cos\phi_0$ dependence  on the phase factor in (\ref{mean_torque_11}), the mean effect being maximal when $\phi_0 =0$, i.e.~when $D(\tau)$ and $\omega(\tau)$ are in phase. 

If $\phi_0=\pi/2$ the disc motion is reversed from one half-period to the next.  By the reversibility theorem for Stokes flow, the fluid flow is similarly reversed, and the contributions to the mean torque are therefore equal and opposite in the two half-periods.  This explains why a mean torque is generated only if $\phi_0\ne \pi/2$.

\begin{figure}
        \centering
        \begin{minipage}{.33\textwidth}
            \centering
            \includegraphics[width=.9\linewidth]{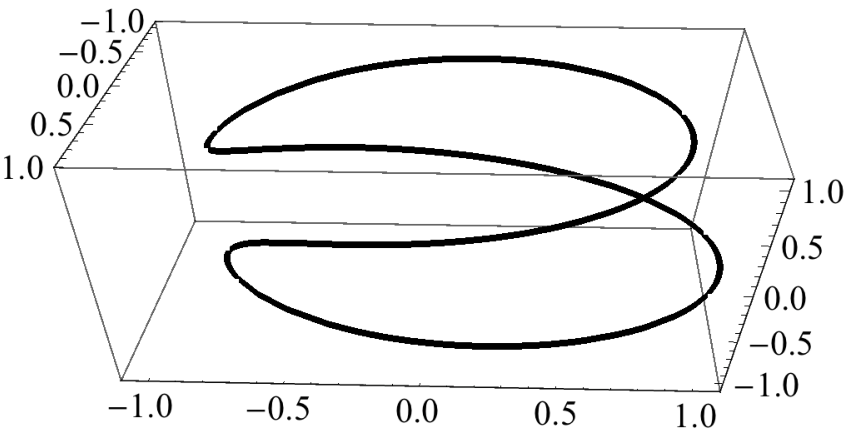}
            \vspace{-4mm}
            \caption*{(a) $n=1$}
        \end{minipage}%
        \begin{minipage}{.33\textwidth}
            \centering
            \includegraphics[width=.9\linewidth]{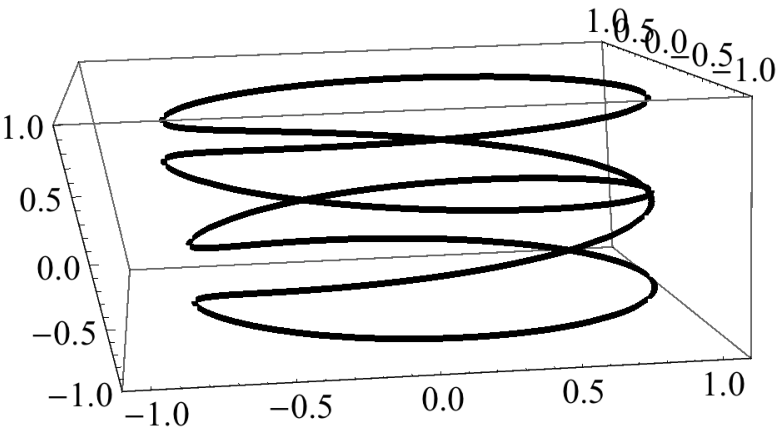}
            \vspace{-4mm}
            \caption*{(b) $n=2$}
        \end{minipage}
        \begin{minipage}{.33\textwidth}
            \centering
            \includegraphics[width=.9\linewidth]{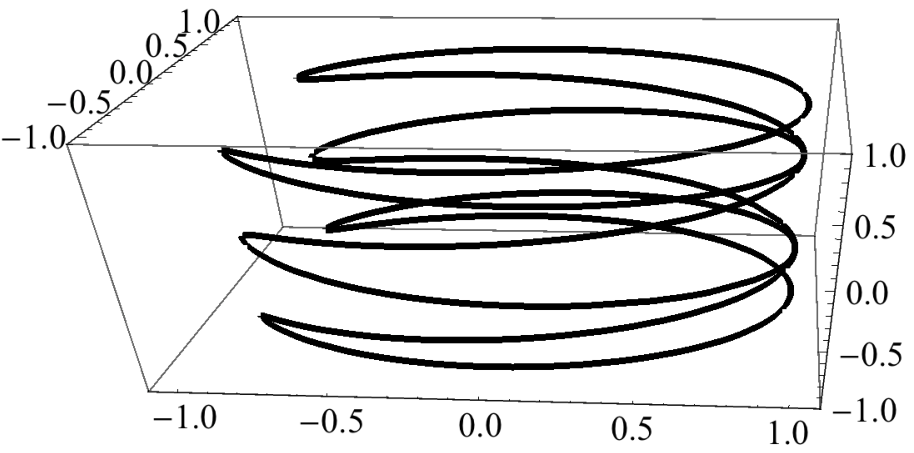}
            \vspace{-4mm}
           \caption*{(c) $n=3$}
        \end{minipage}
        \vskip -2mm
\caption{Trajectories of a point on the disc with velocity (\ref{reg2}) for different $n$;  $d_0 =0, \,\zeta_0=.8$, \,$\omega_0=2.5$.}
\label{Fig_traj_centre}
    \end{figure}

\begin{figure}
        \centering
        \begin{minipage}{.33\textwidth}
            \centering
            \includegraphics[width=.9\linewidth]{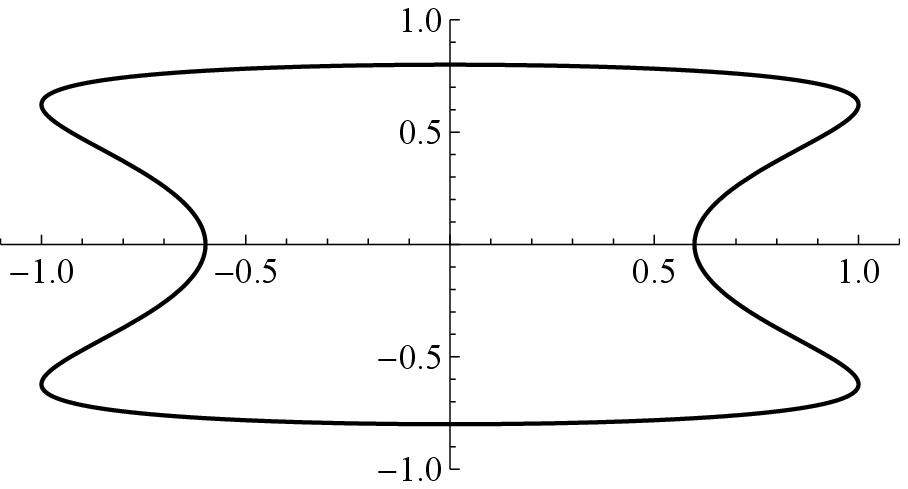}
            \vspace{-4mm}
            \caption*{(a) $n=1$}
        \end{minipage}%
        \begin{minipage}{.33\textwidth}
            \centering
            \includegraphics[width=.9\linewidth]{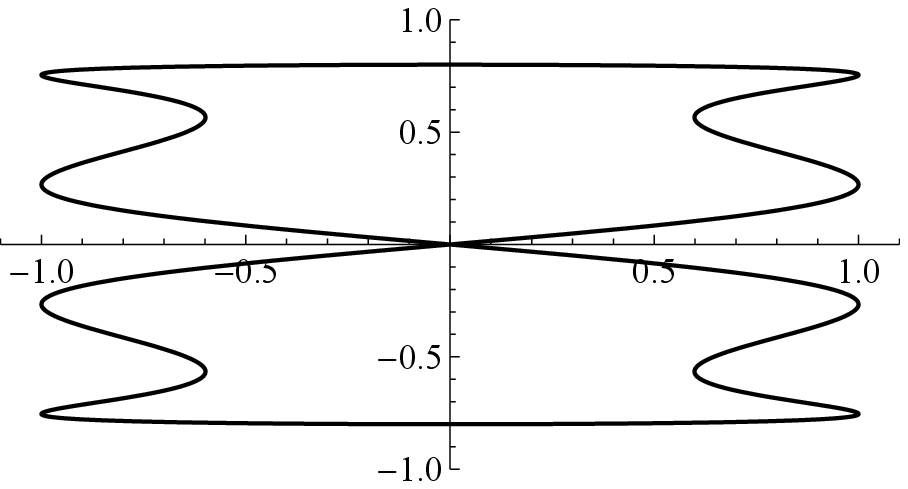}
            \vspace{-4mm}
            \caption*{(b) $n=2$}
        \end{minipage}
        \begin{minipage}{.33\textwidth}
            \centering
            \includegraphics[width=.9\linewidth]{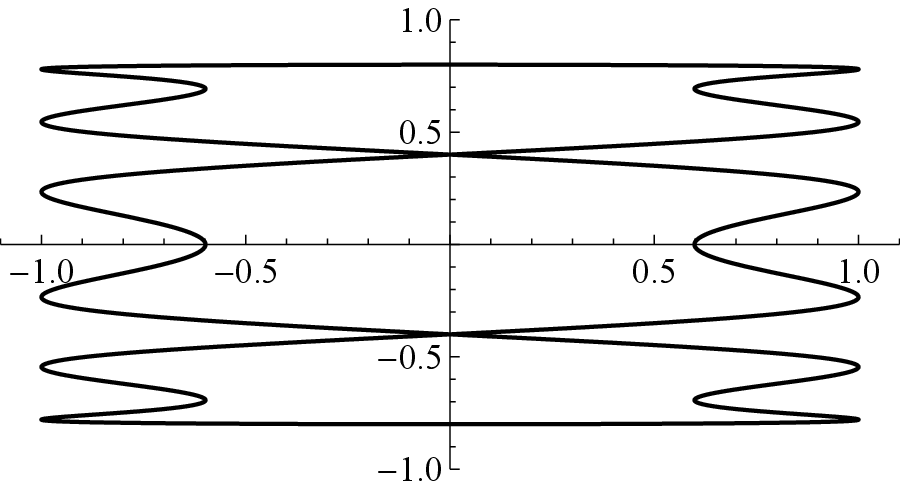}
            \vspace{-4mm}
           \caption*{(c) $n=3$}
        \end{minipage}
        \vskip -2mm
\caption{Projections on the plane $x=0$ of the trajectories of Figure \ref{Fig_traj_centre}.}
\label{Fig_proj_centre}
    \end{figure}

When  $d_0=0$ in \eqref{reg1}, then in \eqref{mean_torque_1}  $\left<G\right>=0$ also, i.e.~the total averaged torque generated on both boundaries taken together is zero, as might be expected from symmetry.  The above physical description suggests however how torque may be generated through modification of $\omega(\tau)$.  Thus suppose for example that
\begin{equation}\label{reg2}
  \omega(\tau)=\omega_{0}\cos n\tau, \quad d(\tau)=\zeta_0\cos\tau,
\end{equation}
where $n$ is a positive integer. 
Figure \ref{Fig_traj_centre}  shows three possibilities for the trajectories of a point on the disc,  and Figure \ref{Fig_proj_centre} shows corresponding projections with $\omega(t)/\omega_0= \cos \tau, \,2 \cos 2\tau$, and $3 \cos 3\tau$ (so $\varphi(\tau)/ \omega_0 =\sin\tau,\,  \sin 2\tau, \,\sin 3\tau$ respectively). 
%\textcolor[rgb]{1.00,0.00,0.00}{COMMENT: explanation on the mening of that pictures should be installed here.}
For $n=1$ or $3$ (and more generally for any odd $n$) such a trajectory has the opposite sense (clockwise/anticlockwise) near the boundaries $z=\pm 1$, whereas for $n=2$  (and more generally for any even $n$) it has the same sense near both boundaries.  We may then expect  that $\langle G\rangle$ should be nonzero only if $n$ is even.

\begin{figure}
\vskip 5mm
\begin{center}
\hspace*{0pt}
\includegraphics[width=.5\textwidth, trim=0mm 0mm 0mm 0mm]{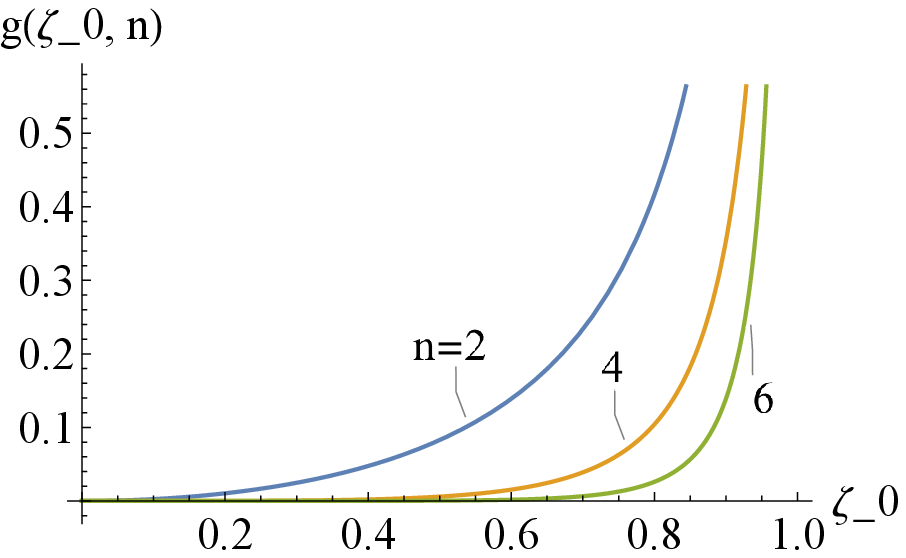}
\end{center}
\vskip 2mm
\caption{ The functions $\langle g(\zeta_0,n)\rangle$  for $n=2,4,6$, defined by (\ref{symm_2}).}
\label{Fig_gzeta}
\end{figure}

This is confirmed by evaluation of \eqref{G1} for the motion \eqref{reg2}, which yields (with the help of {Gradshtein \& Ryzhik 2007}, p.391, formula 3.613)
\be\label{symm_2}
\langle G(\zeta_0,n)\rangle=\kappa\,\omega_0\, g(\zeta_0,n),
\ee
where
\be
g(\zeta_0,n)=\frac{1+(-1)^n}{2\sqrt{1-\zeta_0^2}}\left(\frac{1-\sqrt{1-\zeta_0^2}}{\zeta_0}\right)^{\!\!n}\,.
\ee
\!For $n$ even, $g(\zeta_0,n)\sim (\zeta_0/2)^{n}$ for small $\zeta_0$; the first three ($n=2,4,6$) are plotted in Figure \ref{Fig_gzeta}.  Note again the (expected) singular behaviour as $\zeta_0\rightarrow 1$, when the disc makes instantaneous contact with each boundary $z=\pm 1$ once in each period of the motion.

The quantity $\chi=\left<\omega(\tau)D(\tau)\right>$ that appears in the above examples
is a pseudo-scalar, which can be non-zero only if the motion of the disc is chiral in character 
(i.e.~lacking reflection symmetry).
It is in this respect that the phenomenon of mean-torque generation is analogous to the phenomenon of spontaneous dynamo-generation of magnetic field in a conducting fluid, resulting from non-zero mean helicity of a random wave field.  In that context, a phase shift between velocity and magnetic perturbations is required to provide a nonzero $\alpha$-effect (Moffatt 1978, \S 7.7).  Similarly, in the present context,  there must be a phase shift  between the vertical displacement $d(\tau)$ of the disc and its angular velocity $\omega (\tau)$ to provide a nonzero mean torque.
The pseudo-scalar $\chi$  is nonzero for both cases \eqref{reg1} and  (\ref{reg2} with $n$ even)
and  provides the appropriate measure of chirality of the disc motion.

\section{The two-sphere problem}
Similar effects are to be expected for the two-sphere problem illustrated in Figure \ref{Fig_config}(a).
Here we shall assume that the outer sphere $S_2$ with radius $r_2$ is fixed, and that the inner sphere $S_1$ with radius $r_1 < r_2$ oscillates on the line of centres so that the separation between the centres is 
\be
d(t)=d_0+\zeta_0\cos \tau\quad \textnormal{with}\,\,\,\, 0<d_0\pm\zeta_0 <r_2-r_1\,\,\,\,\textnormal{and}\,\,\,\, \tau=\sigma\, t, 
\ee
and 
simultaneously about the line of centres with time-periodic angular velocity $\omega(\tau)$, with period $2\pi$ and $\left<\omega(\tau)\right>=0$. The dotted spheres in Figure \ref{Fig_config}(a) indicate the range of the up-down oscillation.  We further assume that suitably defined Reynolds and Stokes numbers are small, so that the flow, governed by the Stokes equations, is quasi-static and instantaneously determined by the no-slip boundary conditions. To this extent, the formulation is very similar to that of \S\ref{Sec_cartesian}.

\begin{figure}
\vskip 0mm
\begin{center}
\begin{minipage}{0.99\textwidth}
%height=0.187\textheight
%height=0.165\textheight
\hspace*{0pt}
(a)\,\,\includegraphics[width=0.35\textwidth, trim=0mm 0mm 0mm 0mm]{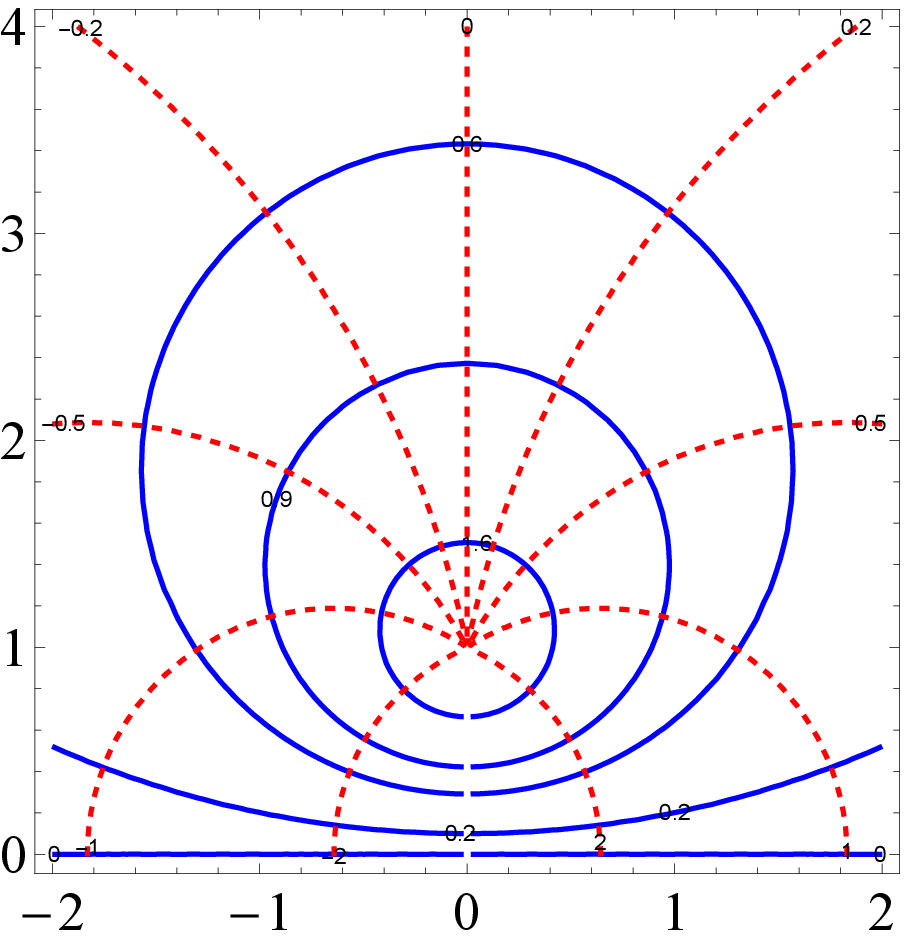}
\hspace*{20pt}
(b)\includegraphics[width=0.45\textwidth,  trim=0mm 0mm 0mm 0mm]{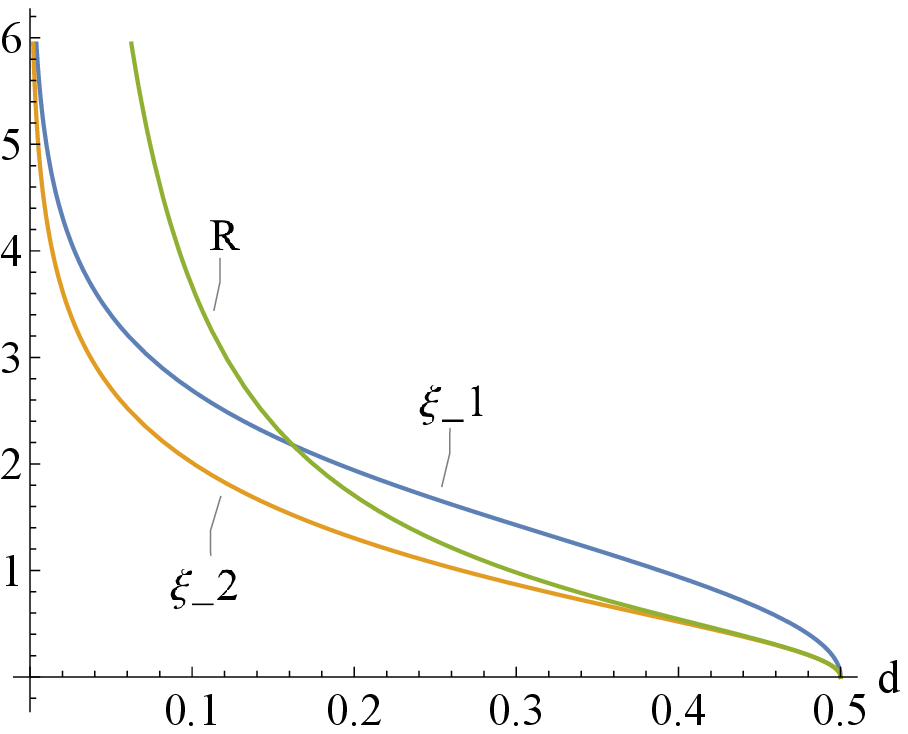}
\end{minipage}
\end{center}
\vskip 2mm
\caption{(a) Contours $\xi=$ const. (solid)  and $\eta=$ const. (dashed) in the half-plane $z>0$, as given by (\ref{complex}) for $R=1$; (b) the functions $\xi_{1}(d)$,  $\xi_{2}(d)$ (which tend logarithmically to infinity as $d\rightarrow 0$) and $R(d)$ for $r_1 =0.5,\,r_2 =1$, as given by (\ref{ks12})  and  (\ref{R}).} 
\label{Fig_contours}
\end{figure}

Following Jeffery(1912, 1915) and Papavasiliou \& Alexander (2017), we adopt bi-spherical polar coordinates $(\xi,\varphi, \eta)$ defined in terms of cylindrical polar coordinates $(r, \varphi, z)$ by
\begin{equation}\label{complex}
  \xi+\textnormal{i}\,\eta=\log\left[\frac{r+\textnormal{i}\,(z+R)}{r+\textnormal{i}\,(z-R)}\right]\quad \textnormal{or equivalently}\quad r+\textnormal{i}\,z=R\,\frac{\sin\eta+\textnormal{i}\sinh\xi}{\cosh\xi-\cos\eta}\,,
\end{equation}
where $R$ is an arbitrary positive real number. The contours $\xi=$ const. (solid) and $\eta=$ const. (dashed) are  shown in Figure \ref{Fig_contours}(a).  From (\ref{complex}), it may be ascertained that
\be
r^2 +(z-R\,\textnormal{coth}\, \xi)^2=R^{2} \textnormal{cosech}^{2}\xi\,.
\ee
Thus the contours $\xi=$ const. are circles with centres at $r=0, \,z=R\,\textnormal{coth}\, \xi$, and radii $R\, \textnormal{cosech}\,\xi$.  For given $(r_1,\,r_2,\,d)$, it follows that $\xi_1,\,\xi_2$ and $R$ satisfy
\be\label{R}
R=r_1 \sinh \xi_1 =r_2\sinh\xi_2,
\ee
\noindent and
\be\label{solve_xi}
r_1\sinh\xi_1-r_2\sinh\xi_2=0, \quad r_1\cosh\xi_1-r_2\cosh\xi_2=-d.
\ee
These equations may be solved for $\cosh\xi_1$, $\cosh\xi_2$ and $R^2$ in terms of $d$, giving
\be\label{ks12}
\cosh\xi_1=\frac{{r_2^{2}-r_1^{2}}-d^2}{2r_1 d}\,,\quad
\cosh\xi_2=\frac{{r_2^{2}-r_1^{2}}+d^2}{2r_2 d}\,,
\ee
and
\be\label{R}
R^2=\left((r_1-r_2)^2-d^2\right)\left((r_1+r_2)^2-d^2\right)/4d^2\,.
\ee
\!Note that $d$ must be non-negative for real $\{\xi_1,\xi_2\}$. We may take  $0\! <\! d\! <\! r_{2}-r_{1}$; these inequalities ensure that $\xi_1\! >\! \xi_2\! >\! 0$.   Figure \ref{Fig_config}(a) actually shows the situation when $r_1=0.5, \,r_2=1,\,d=0.25$ and $\zeta_{0}=0.2$, the dotted spheres indicating the limiting positions in the up-down oscillations. Fig \ref{Fig_contours}(b) shows the functions $\xi_1 (d)$ and $\xi_2 (d)$ when $r_1=0.5,\,r_2=1$.
More generally, $\xi_1 (d)$ and $\xi_2 (d)$ both vanish when $d=r_2-r_1$, i.e.~when the spheres are in contact.  From (\ref{R}), we then have asymptotically
\be
r_1\xi_1\sim r_2\,\xi_2 \quad\textnormal{as}\,\,d\rightarrow r_2-r_1.
\ee
In this contact limit, one might expect the torque to be dominated by local conditions near the contact point, but this is not in fact the case (see Appendix).
\section{ Transfer of angular momentum}
\begin{figure}
\begin{center}
\vskip 0mm
%height=0.187\textheight
%height=0.165\textheight
\hspace*{0pt}
\includegraphics[width=0.50\textwidth, trim=0mm 0mm 0mm 0mm]{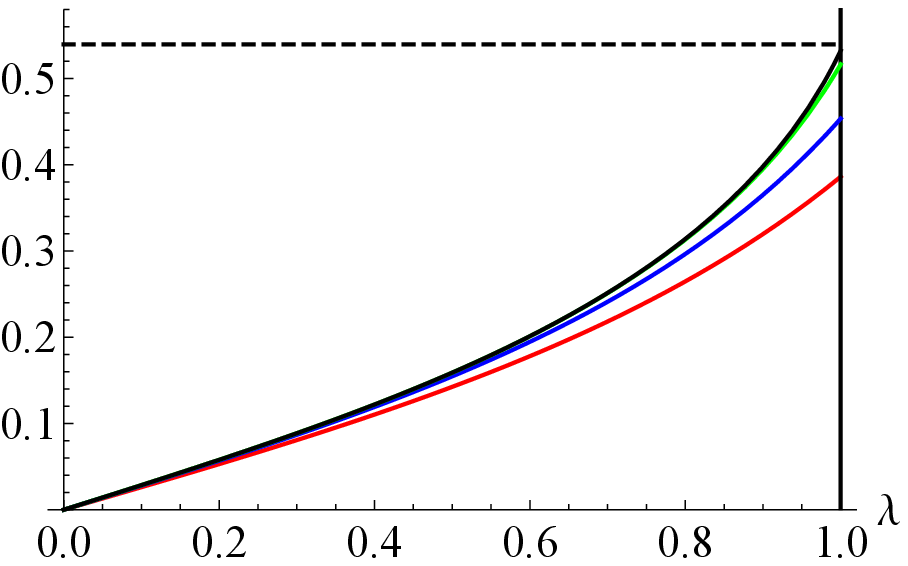}
\end{center}
\vskip 2mm
\caption{  $\sum_{m=1}^{n} g_{m}(\lambda)$ for $r_1=0.5, d_{0}=0.25$ and for $n=2$ (red), 3 (blue), 6 (green) and 10 (black); the sum to infinity is uniformly convergent, and in particular $\sum_{m=1}^{\infty} g_{m}(1)\approx 0.539667$.} 
\label{Fig_g}
\end{figure}
In the Stokes approximation, 
with a purely azimuthal velocity field ${\bf u}=(0, v(r,\,z),0)$,  
the pressure is constant throughout the fluid and 
\be\label{stokes}
(\nabla^2-r^{-2})v=0.
\ee
The general solution of this equation in the above bi-spherical coordinates is:
\begin{equation}\label{gen-sol}
  v=\sqrt{W}\sum_{n=0}^\infty \{a_n \exp[(n+1/2)(\xi-\xi_1)]+b_n \exp[-(n+1/2)(\xi-\xi_2)]\}P_n^1(\cos\eta)\,,
\end{equation}
where $W\equiv R/(\cosh \xi-\cos\eta)$ (Jeffery 1915). The constants $a_n$ and $b_n$ are determined  via the boundary conditions
\begin{equation}\label{bound-cond}
 v(\xi_1)=\omega(\tau) \,r_{1}\,,\quad  v(\xi_2)=0.
\end{equation}
The instantaneous torque $G(\tau)$ exerted on the outer sphere is then given by equation (11) from Jeffery (1915), which may here be expressed in the form
\be
G(\tau)=8\pi\mu r_1^3\omega(\tau) \sum_{m=0}^{\infty}\frac{\sinh^3 \xi_1}{\sinh^3 [(m+1)\xi_1(\tau)-m\xi_2(\tau))]}\,,
\ee\label{inst_torque}
\!\!and the instantaneous torque experienced by the inner sphere is $-G(\tau)$. Since each term of the sum (\ref{inst_torque}) is positive, it follows that $G(\tau)$ has the same sign as $\omega(\tau)$ at each instant $\tau$, in accord with physical intuition\footnote{Jeffery's expression for his $G_1$ is actually the torque exerted by the sphere $S_1$ on the fluid, and not as he states the torque experienced by $S_1$. Of course, in our time-dependent situation, a torque $+G(\tau)$ would have to be exerted on the sphere $S_1$ by some internal battery mechanism to maintain its postulated rotational motion. An internal battery would also be needed to maintain the periodic up-down motion.}.

We are again concerned with the mean torque exerted on the outer sphere, 
\be
\left<G\right>=8\pi\mu r_1^3\sum_{m=0}^{\infty}\left<\frac{\omega(\tau)\sinh^3 \xi_1}{\sinh^3 [(m+1)\xi_1(\tau)-m\,\xi_2(\tau)]}\right>\,.
\ee\label{mean_torque}
If we define $\lambda=\zeta_{0}/d_0$ and take $\omega(\tau)=\omega_{0}\cos(\tau+\psi)$, this becomes
\be\label{mean_torque2}
\left<G\right>(r_1, d_0,\lambda)=4\mu r_1^{3}\omega_{0} \cos\psi\sum_{m=0}^{\infty}g_{m}(r_1, d_0,\lambda)\,,
\ee
where
\be
g_{m}(r_1, d_0,\lambda)=\int_{\!\Delta}\frac{\cos\tau \sinh^3 \xi_1(\tau)}{\sinh^3 [(m+1)\xi_1(\tau)-m\,\xi_2(\tau))]}\,\textnormal{d}\tau\,,
\ee
and $\Delta$ is any $2\pi$-period of $\tau$, and  $\xi_1(\tau)$ and $\xi_2(\tau)$ are given by (\ref{ks12}) with now $d(t)=d_0(1+\lambda \cos \tau)$. Figure \ref{Fig_g} shows the functions $\sum_{m=1}^{n}g_{m}(r_1, d_0,\lambda)$ ($n=2,3,6,10$) with (by way of illustration) $r_1=0.5, d_{0}=0.25$, and for the relevant range of $\lambda$, i.e.~$0<\lambda<1$.  The sum (\ref{mean_torque2}) is uniformly convergent, even at the limiting value $\lambda=1$ for which the spheres make contact and for which $\sum_{m=1}^{\infty} g_{m}(1)\approx 0.539667$. 

The situation here is to be contrasted with that of Figure \ref{Fig_G} which showed a divergent torque when the amplitude of the oscillation was maximal.  In the present spherical case, the influence of the inner sphere on the outer sphere is much weaker in the contact limit, so weak in fact that the resulting torque is finite in this limit.

\section{Oscillations symmetric about the centre of $S_2$}\label{Sec_centred_oscillations}
\begin{figure}
\vskip 5mm
\begin{center}
\hspace*{0pt}
\includegraphics[width=.8\textwidth, trim=0mm 0mm 0mm 0mm]{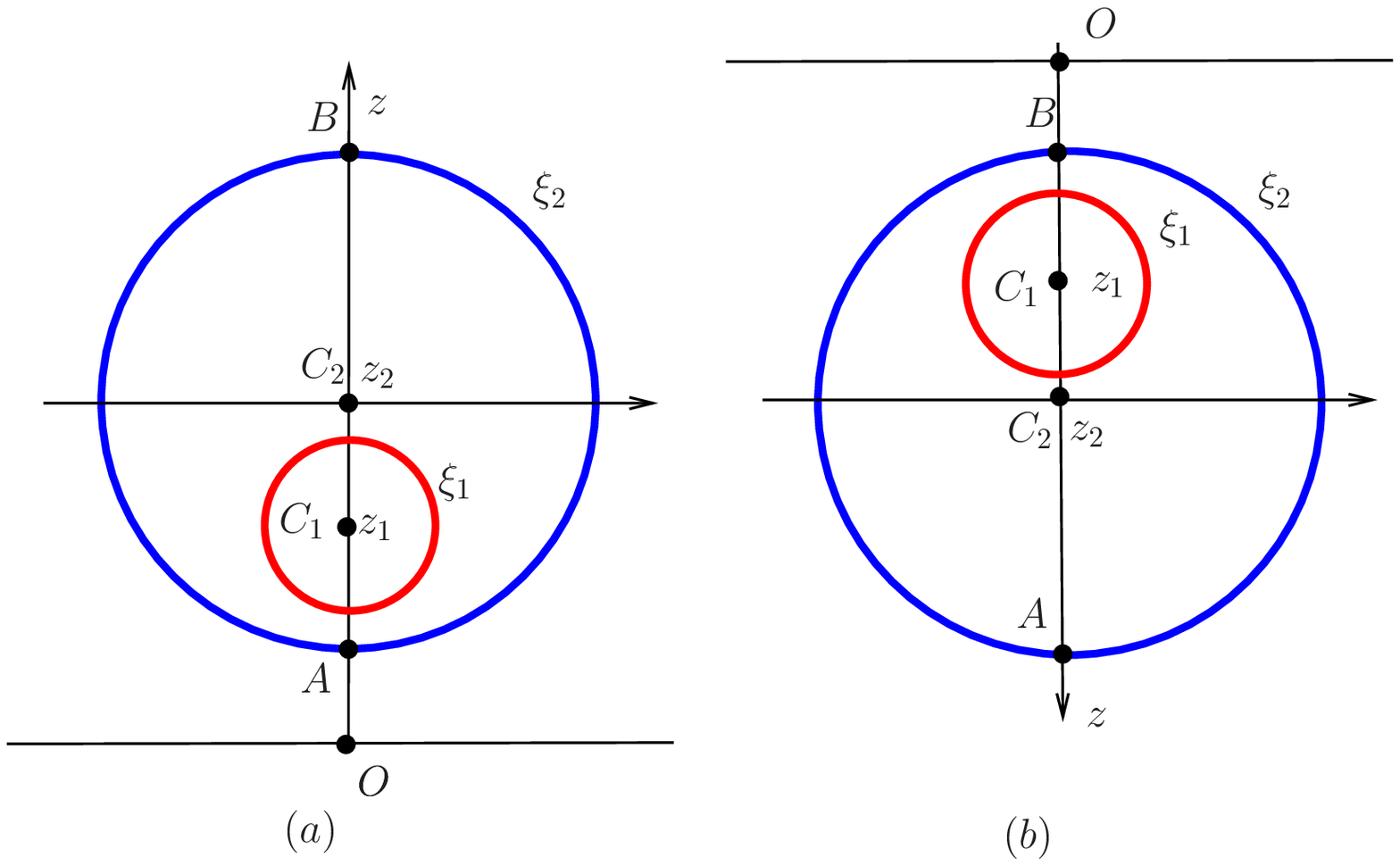}
\end{center}
\caption{ Reversal of $z$-coordinate if $C_1$ passes through $C_2$; (a) $C_1$ below $C_2$; (b) $C_1$ above $C_2$.}
\label{Fig_spheres_inverted}
\end{figure}
The above treatment requires minor modification if $d(\tau)$ is such that the centre $C_1$ of $S_1$ rises above the centre $C_2$ of $S_2$ in the course of its oscillations. As already indicated, $\xi_1$ and $\xi_2$, as defined by (\ref{ks12})  are real only if $d(\tau)>0$ for all $\tau$, i.e.~only if $C_1$ remains below $C_2$.  If  $C_1$ rises above $C_2$, it is necessary to `switch' the coordinate system by simply replacing $z$ by $-z$ in (\ref{complex}); the situation is as represented in the sketch of Figure \ref{Fig_spheres_inverted}. This artifice ensures that $d(\tau)=z_{2}(\tau)-z_{1}(\tau)$ is indeed always positive.

\begin{figure}
        \centering
        \begin{minipage}{.4\textwidth}
            \centering
           (a) \includegraphics[width=.9\linewidth]{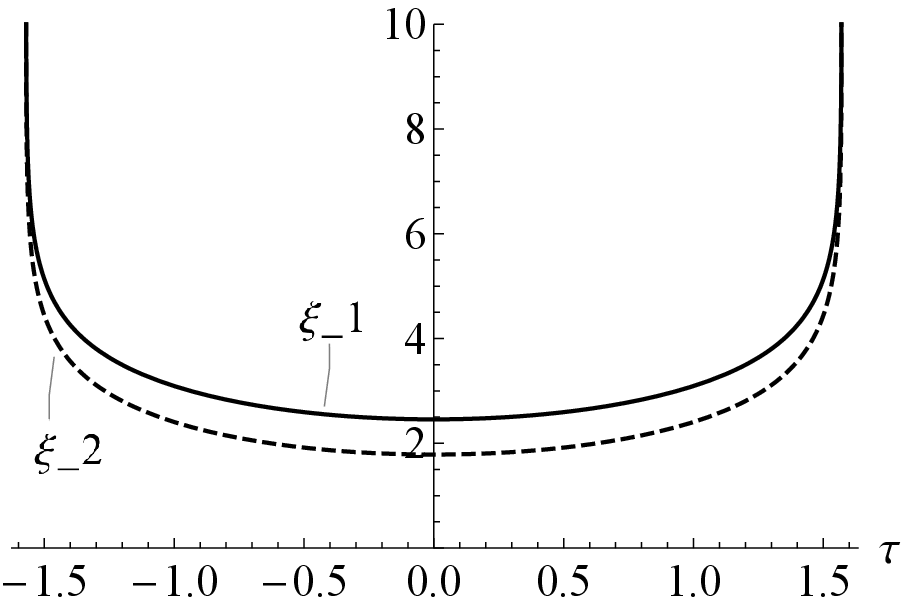}
        \end{minipage}%
        \hskip 3mm
        \begin{minipage}{.4\textwidth}
            \centering
            (b)\,\,\includegraphics[width=.9\linewidth]{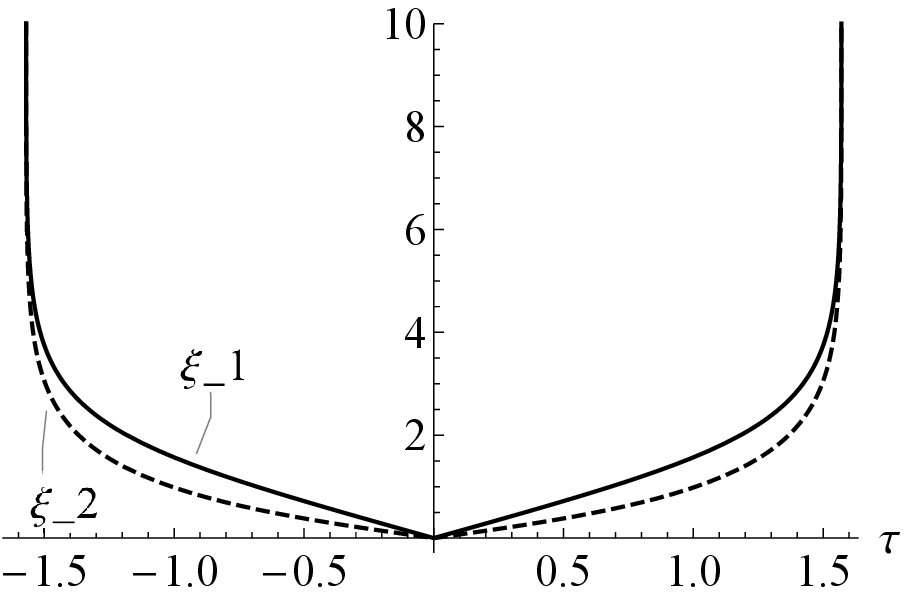}
        \end{minipage}
        \caption{ $\xi_{1}(\tau)$ and $\xi_{2}(\tau)$ in the time interval $-\pi/2<\tau<\pi/2$ for (a) $\zeta_0=1/8$, (b) $\zeta_0=1/2$.}
        \label{Fig_ksi_centred}
    \end{figure}
\begin{figure}
\vskip 5mm
\begin{center}
\hspace*{0pt}
\includegraphics[width=.5\textwidth, trim=0mm 0mm 0mm 0mm]{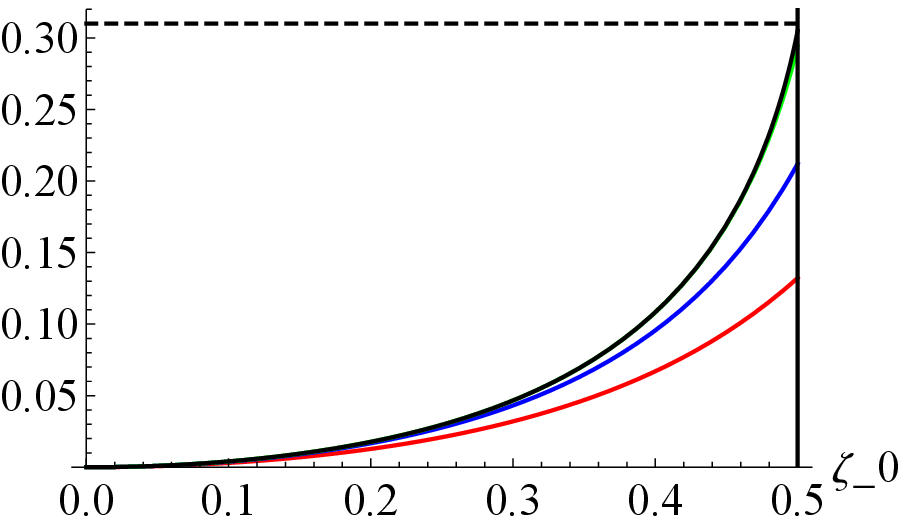}
\end{center}
\vskip 2mm
\caption{The torque $\left<G\right>(\zeta_0)/4\mu r_1^{3}\omega_{0}$ given by (\ref{torque_center}) and  (\ref{torque_center_2}) for $r_{1}=1/2$ and $\omega=\omega_{0}\cos 2\tau$; the partial sums $\sum_{m=1}^{n}h_{m}(\zeta_0)$ are shown for $n=1$ (red), 2 (blue), 6 (green) and $10$ (black); the sum converges to 0.3099  at $\zeta_0=1/2$ when the torque is maximal.}
\label{Fig_h}
\end{figure}
Suppose for example that $d(\tau)$ is initially specified as $d(\tau)=\zeta_0 \cos\tau$, where the amplitude $\zeta_0$ satisfies $0<\zeta_0<r_{2}-r_{1}$.  This may be used only when  $d(\tau)>0$, for example in the time interval $-\pi/2<\tau<\pi/2$. Figure \ref{Fig_ksi_centred} shows $\xi_{1}(\tau)$ and $\xi_{2}(\tau)$ in this interval for the particular choices $\zeta_0=1/8$ and $\zeta_0=1/2$; the latter choice is maximal -- for this value, the two spheres touch at time $\tau=0$.   Now if $\omega=\omega_{0}\cos n\tau$, then just as in \S 2, the mean torque $\left<G\right>$ on the sphere $S_2$ is zero or non-zero according as $n$ is odd or even. Moreover, in the latter case, symmetry (in the mean) about the horizontal plane through $C_2$ allows us to calculate $\left<G\right>$ by simply averaging over the $\pi$-interval $-\pi/2<\tau<\pi/2$ (since the mean torque over the subsequent $\pi$-interval $\pi/2<\tau<3\pi/2$ will obviously be the same). Just as in (\ref{mean_torque2}), with $\omega=\omega_{0}\cos 2\tau$ the mean torque is therefore given by
\be
\left<G\right>(r_1, \zeta_0)=4\mu r_1^{3}\omega_{0}\sum_{m=0}^{\infty}h_{m}(r_1, \zeta_0)\,,
\label{torque_center}
\ee
where now
\be
h_{m}(r_1,\zeta_0)=\int_{-\pi/2}^{\pi/2}\frac{\cos 2\tau \sinh^3 \xi_1(\tau)}{\sinh^3 [(m+1)\xi_1(\tau)-m\,\xi_2(\tau))]}\,\textnormal{d}\tau\,.
\label{torque_center_2}
\ee
Figure \ref{Fig_h} (which may be compared with  Figure \ref{Fig_g}) shows the partial sums $\sum_{m=1}^{n}h_{m}(\zeta_0)$ for $n=1$ (red), 2 (blue), 6 (green) and 10 (black).  The convergence is obviously uniform and rapid.  The sum converges to 0.3099  at $\zeta_0=1/2$ when the torque is maximal. Again the contrast between Figure \ref{Fig_g} and Figure \ref{Fig_gzeta} which showed a divergent torque should be noted.  For the spherical geometry, the interaction between the spheres is again so weak in the contact limit that the transmitted torque remains finite in this limit.

The above technique may be easily adapted to cover the more general case when 
$0\le d_0 <\thalf(r_{2}-r_{1}),\, d_0<\zeta_{0} < r_{2}-r_{1}-d_0$, when the amplitude of the oscillations  is large enough for $C_1$ to rise above $C_2$ but still small enough for $S_1$ to remain inside $S_2$ throughout the period of the oscillation.  We need not labour the details here.

\section{Conclusions and discussion}
For the cartesian geometry of Figure \ref{Fig_config}(b), we have shown by straightforward analysis
that time-periodic axisymmetric motion of the disc  with zero mean can generate a mean torque on the two fixed plates bounding the fluid domain; in this, we have assumed quasi-static Stokes flow in which all inertia effects are negligible. Mean torque generation requires that the disc motion should have a chiral character, as measured by the pseudo-scalar quantity $\chi=\langle \omega(\tau) D(\tau)\rangle$, where $D(\tau)=(1-d(\tau)^2)^{-1}$, and also that there should be a phase shift between its vertical and rotational components. In these respects, the generation of a mean torque is analogous to the dynamo generation of a mean (large-scale) magnetic field by a random-wave turbulence that is chiral and for which a phase shift is generated between the velocity and the resulting magnetic perturbation. 

For the spherical geometry of  Figure \ref{Fig_config}(a), we have demonstrated a similar phenomenon when the outer sphere $S_2$ is fixed and the inner sphere $S_1$ is subject to a time-periodic motion with zero mean velocity.  We have given a clear physical interpretation of the effect, namely that when $S_1$ is near to $S_2$ the rotational torque transmitted to $S_2$  is stronger than when it is far from $S_2$.  The phase shift is required, because without it the disc motion is time-reversible, and the reversibility theorem for Stokes flow ensures that, although the disc motion is still chiral, the instantaneous torque generated necessarily averages to zero. The spherical problem is significantly different from the planar problem in that the mean torque transmitted remains finite even in the limit when the spheres make instantaneous contact during a period of the up-down oscillation.

Although the analysis of the paper is restricted to Stokes flow, there seems little doubt that the effect must persist when fluid inertia is taken into account, because the above physical explanation is still applicable. It would be reasonably straightforward to take account of increasing frequency of the disc motion (i.e.~increase of Stokes number), because the problem then remains linear.  It would be much more difficult to take account of increasing Reynolds number, because there is then a nonlinear interaction between the poloidal and toroidal ingredients of the flow.  Again, the problem has some similarity with the dynamo problem, in that explicit calculation of the $\alpha$-effect in dynamo theory is easy only when the magnetic Reynolds number $Re_m$ based on the random-wave scale is small, and very difficult when $Re_m\gg 1$ (Moffatt \& Dormy 2019).

There is moreover no need to restrict the boundaries  $S_1$ and $S_2$ to be spherical; the effect of chiral transfer will still obviously occur under similar circumstances if they are axisymmetric with a common axis of symmetry; and indeed the general principle of chiral transfer of angular momentum may be expected to have wide generality.

\oneappendix
\appendix

\section{Comment on the contact  limit}
\begin{figure}
\vskip 5mm
\begin{center}
\hspace*{0pt}
\includegraphics[width=.3\textwidth, trim=0mm 0mm 0mm 0mm]{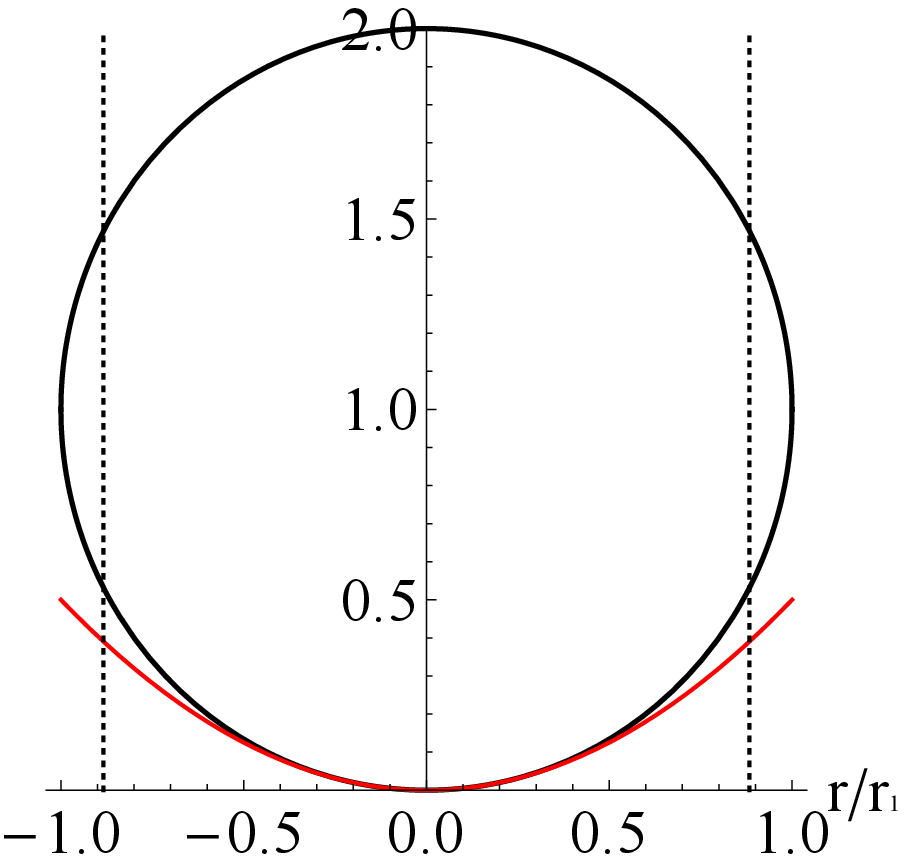}
\end{center}
\vskip 2mm
\caption{The limit when the sphere $S_1$ touches the plane $z=0$;  the sphere may be approximated near $r=0$ by the paraboloid as shown in red, but the dominant contribution to the total torque comes from the region indicated by the dotted lines where lubrication theory is not applicable.}
\label{Fig_lubrication}
\end{figure}

We comment here on the inapplicability of conventional lubrication theory when the minimum gap between the two spheres tends to zero.  It is sufficient to consider just the situation when $r_2\rightarrow\infty$ so that $S_2$ becomes the plane boundary $z=0$,  and when the sphere $S_1$ makes contact with this plane; the situation is shown in Figure \ref{Fig_lubrication}.  If in this situation $S_1$ rotates about the diameter through the point of contact with {\em steady} angular velocity $\omega_0$, then, as observed by Papavasiliou \& Alexander (2017),  the torque $G$ given by (\ref{inst_torque}) reduces to
\be\label{inst_torque_limit}
G\sim G_0=-8\pi\,\mu\, r_{1}^3\, \omega_0\, \zeta (3),
\ee
where $\zeta(s)=\sum_{n=1}^{\infty}n^{-s}$ is the Riemann zeta function. When the sphere $S_1$ is remote from the plane, the well-known expression for the torque is
\be
G\sim G_{\infty}= -8\pi\,\mu\, r_{1}^3\, \omega_0,
\ee
so that
\be \label{torque_zero}
G_0/G_{\infty}=\zeta (3)\approx 1.20205.
\ee
Jeffery (1915) actually tabulated the value of the torque for various values of the ratio of the radius $r_1$ to the distance $r_1 +\epsilon$ where $\epsilon$ is the minimum gap between the sphere and the plane, and noted that, if  $\epsilon/r_1=0.02$, ``the couple required to maintain the rotation is only increased [from $G_{\infty}$] by 17\%". We see from the above that in the contact limit $\epsilon=0$, the couple is increased by only 20.2\%.

Conventional lubrication theory (Batchelor 1967), would approximate the sphere $z/r_1=1-(1-(r/r_{1})^{2})^{1/2}$ near the point of contact by the paraboloid
\be
z/r_1=h(r)=\thalf (r/r_1)^{2},
\ee
shown in red in Figure \ref {Fig_lubrication}. Lubrication theory is justifiable only in the `contact region' $r\ll r_1$. In this small region, the velocity profile is linear, i.e.~locally Couette, and the contribution to the torque on $S_1$ out to radius $r$ in this region is 
easily calculated as
\be
G(r)\sim -2\pi \,\mu\,\omega_{0}\int_{0}^{r} \frac{r'^{3}}{h(r')}\textnormal{d}r' =  -2\pi\,\mu r_{1} \omega_0\, r^{2}\,.
\ee
\!This increase with $r$ resulting from the factor $r'^3$ in the integrand, indicates that the dominant contribution to the total torque in fact comes from outside this contact region, i.e.~from the region $r=O(r_1)$  indicated by the dotted lines in Figure \ref {Fig_lubrication}; there can be no justification for use of the lubrication approximation throughout this extended region.

\begin{acknowledgments}
\noindent We acknowledge helpful discussion with Prof. Gareth Alexander, University of Warwick.
This research is partially supported by grant IG/SCI/ DOMS/18/16 from the Sultan Qaboos University, Oman.
\end{acknowledgments}

\end{document}